\documentclass[11pt]{article}
\usepackage{moriond}
\usepackage{amsfonts}
\def\Journal#1#2#3#4{{#1} {\bf #2}, #3 (#4)}

\def\PLB{{\em Phys. Lett.}  B}

\def\APP{{\em Acta Phys. Polon.} B}
\def\PRS{{\em Proc. R. Soc. London} A}
\def\CMP{\em Commun. Math. Phys.}
\def\NLN{\em Nonlinearity}
\newcommand{\e}{\mathbf{e}}
\newcommand{\f}{\mathbf{f}_{\pi}}

\newcommand{\Tr}{\textsl{Tr}}
\newcommand{\ud }{\mathrm{d}}
\newcommand{\ha}[1]{\mathbf{H}_{\scriptscriptstyle{{#1}}}}
\newcommand{\sk}[1]{\mathbf{S}_{\scriptscriptstyle{{#1}}}}

\begin{document}
\vspace*{4cm}
\title{UNDERSTANDING THE 1-SKYRMION}
\author{ \L. BRATEK }
\address{Institute of Nuclear Physics,
  Polish Academy of Sciences,\\
  ul. Niewodniczanskiego 152,
  31-242 Krakow,
  Poland,\\
  e-mail: lukasz.bratek@ifj.edu.pl}

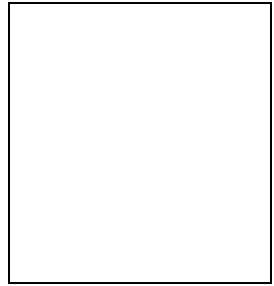
\maketitle\abstracts{It is shown how the appearance of the
$1$-skyrmion on the three-sphere of radius $L$ can be understood
quantitatively by analyzing spectrum of the Hessian at the
identity solution. Previously, this analysis was done
qualitatively by N. Manton \cite{manton87} which used a conformal
deformation of the identity solution. Analysis of critical
appearance of solutions in the Skyrme model on the three-sphere
may be useful to understand similar phenomena in more complicated
equations of mathematical physics. }

\section{Some General Remarks}

In this paper static solutions of the $SU(2)$ Skyrme model on the
three-sphere of radius $R$ as a physical space are discussed. In a
sense, this model includes the Skyrme model on the flat space
$\mathbb{R}^3$ in the limit $R\rightarrow \infty$.

Historically, the Skyrme model was introduced to physics as a
mesonic fluid model describing nuclear matter, and was proposed by
T.H.R Skyrme in 1954  \cite{skyrme54}. In agreement with the
Yukawa picture, nuclear interactions are mediated by $\pi$-meson
fields that are regarded as constituents of nucleons. The idea of
Skyrme was to imagine such fields as an incompressible
 mesonic fluid. The state of such a fluid is defined by a scalar
density and a direction in the isospace. Nucleons are
assumed to be immersed into a mesonic liquid droplet,
freely moving inside it forming a nucleus.
In this effective meson field theory,
low-energy baryons emerge as solitonic fields.
The basic chiral field in that model is the $SU(2)$-valued
scalar field $U$, which is related to the singlet meson
field $\sigma $ and the triplet pion field $\vec{\pi }$ by
$$U(\vec{x})=
\f^{-1}(\sigma(\vec{x}) +i\vec{\sigma}\circ \vec{\pi
}(\vec{x})),\qquad (\sigma)^2 +(\vec{\pi })^2=\f ^2.$$ The
model is characterized by two constants: $\f$, the pion
decay constant and $\e$, determining the strength of the
fourth-order term that was introduced by Skyrme in
\cite{skyrme3} to ensure the existence of
solitonic solutions.
In a generally covariant form the model is defined by the
action integral
$$\mathcal{S}[U]=
\int \sqrt{-\mathrm{det}(g_{\mu\nu}})\ud^4x\left(\frac{1}{4}\f^2\Tr\left(C_\mu
C^\mu\right) +\frac{1}{32\e^2}\Tr\left([C_\mu ,C_\nu][C^\mu
,C^\nu]\right)\right) \label{eqeig:lagdens},$$
where  $C_\mu=i U^+\partial _\mu U$
is the Lie algebra valued topological
four-current and $g_{\mu\nu}$ is the metric tensor.

Currently, apart from its successful applications to nuclear and
condense matter physics, the Skyrme model, as an example of a
nonlinear field-theoretical model, is utilized in studies of
Einstein's field equations (for a review of similar topics you may
consult \cite{bizrev}). The Einstein-Skyrme model \cite{bizchmaj}
in the spherically symmetric case, thus more tractable and also
physically most interesting, reduces to a dynamical system which
possesses a very rich structure of solutions. The spectrum of soliton solutions consists of two branches which coalesce when coupling
between the Skyrme's nonlinear matter and gravity is sufficiently
large. For each winding number this happens
at some critical value of the coupling constant which is a
combination of the Newton's constant $G$ and the dimensional
constant $\f$ by which the nonlinear field couples to gravity. In the limit $G\to0$ one obtains, among other solutions, flat space-time skyrmions.  Due to
complex structure of these equations it seems reasonable to
investigate general properties of the system in a simpler setup.
It turns out, that many characteristics of gravitating Skyrmions
may be observed in the Skyrme model on the three sphere.

The key idea is to 'simulate'
gravitational interactions by considering the Skyrme fields in a
given space-time background. For static solutions we
assume that the space-time is time
invariant and has a constant and positive space-like curvature.
Therefore, we assume that the three-sphere of radius $R$
will play the role of our physical space. Its curvature may
be thought as representing a gravitational field whose
intensity is dependent on how strongly this space is
curved. Therefore, we will see how curvature, and so 'gravity',
affects spectra of solutions.

The Skyrme model on the three-sphere contains two
natural scales of length, namely,
the radius $R$ of the base three-sphere and the characteristic
soliton size $(\f\e)^{-1}$, which break scale invariance.
These scales constitute a free dimensionless
parameter whose value is crucial to the number of possible
solutions. These solutions are topological solitons, that is,
localized, finite energy
field configurations with nontrivial topology.

Due to the existence of the free parameter
this model
generates a denumerable and infinite spectrum of real numbers $1/\alpha$
which can be interpreted as critical radii (in units where
$\e\f=1$) of the base three-sphere, below which some solutions
cease to exist. In a sense, this phenomenon is analogous to the
formation of black-hole solutions in the Einstein-Skyrme model.
The base three-sphere may be too small to support some solutions
since curvature may be too high (i.e. 'gravitational' field too
strong). Indeed, playing with $R$, $\e$ and $\f$ we can construct
the 'gravitational constant' $G(R)=1/(R\e\f^2)^2$ (which is
proportional to the Gauss curvature of our three-space), then a
soliton which has the characteristic mass $\f/\e$ can not exist
when its size is comparable with its 'Schwarzschild radius' - the
characteristic solitonic size $1/(\e\f)$. In this special case it
happens when $R\e\f=\sqrt{2}$ which turns out to be of the same rank
(actually this happens accidentally to be exactly the same value)
of the radius of the three-sphere below which the $1$-skyrmion
cease to exist. We also mention the fact, first observed for
gravitating solitons in \cite{bizchmaj}, that the lowest
eigenvalue of the Hessian of the counterpart of the flat space
stable soliton decreases as $\alpha$ grows and tends to zero as
$\alpha$ tends to some critical $\alpha_o$ at which the branch of
gravitating skyrmions coalesces with another branch of unstable
solitons. This catastrophe-theoretical phenomenon is also observed
in our model and, as it will be shown, can be astonishingly simply
tackled analytically.

\section{Some analytical results}

The Skyrme field may be equivalently considered as a map from the
metrical base space, which is the space-time $(\mathcal{M},g)$
(with metric $g$), to the target space $(\mathcal{N},G)$ (with
metric $G$) -- here the $SU(2)$ group. The $SU(2)$ group is the
metrical and topological unit three-sphere. For any map between
metrical manifolds $(\mathcal{M},g)\ni x\rightarrow y(x)\in
(\mathcal{N},G)$ one can construct the Jacobi matrix
$J^a_\alpha=\partial_{\alpha}y^a$ which pulls back the metric $G$
from $\mathcal{N}$ to $\mathcal{M}$ via the mapping
$\hat{G}_{\alpha\beta}=J^a_{\alpha}J^{b}_{\beta}G_{ab}$. The
tensor $\hat{G}$ can be coupled with $g$ according to the
definition $\hat{g}_{\mu\nu}=g_{\mu\nu}+\kappa^2\hat{G}_{\mu\nu}$,
thus $\hat{g}$ may serve as another metric tensor on
$\mathcal{M}$. This provides us with several invariants of
$\hat{G}$ which are obtained as the expansion coefficients of the
function $\mathrm{Det}(A)(\kappa^2)$ with respect to $\kappa^2$,
where $A^{\alpha}_{\beta}=g^{\alpha\gamma}\hat{g}_{\gamma\alpha}$.
In the Skyrme model only the invariants
$g^{\alpha\gamma}\hat{G}_{\gamma\alpha}$ (the so called sigma
term) and $(g^{\alpha\gamma}\hat{G}_{\gamma\alpha})^2-
g^{\alpha\gamma}\hat{G}_{\gamma\delta}g^{\delta\mu}\hat{G}_{\mu\alpha}$
(the Skyrme term) are used to construct the Lagrangian. We impose
on solutions the hedgehog ansatz, that is, for static and
spherically symmetric solutions we require $\Psi=F(\psi)$,
$\Theta=\theta$ and $\Phi=\phi$, where $(\Psi,\Theta,\Phi)$ are
spherical angles on the unit target three-sphere (the manifold of
the $SU(2)$ group), while $(\psi,\eta,\phi)$ are angles on the
base three-sphere of radius $L$. Hence, such solutions are
critical points of the energy functional
\begin{equation}\mathcal{U}[F]=\int \limits_{0}^\pi 4\pi \sin^2{\psi}\ud\psi\bigg\{
L\bigg[F'(\psi)^2+2\frac{\sin^2{F(\psi)}}{\sin^2{\psi}} \bigg]
+\frac{1}{L}\bigg[2F'(\psi)^2+\frac{\sin^2{F(\psi)}}{\sin^2{\psi}}
\bigg] \frac{\sin^2{F(\psi)}}{\sin^2{\psi}}\bigg\},\label{eq:en}\end{equation}
where $L$ is the radius of the base three-sphere, and
the unit of energy and length are respectively $\f \e^{-1}/2$
and $(\e\f)^{-1}$. The ansatz has symmetries of the full Lagrangian
(that is, without any symmetry constraints) thus, according to the principle
of symmetric criticality \cite{palais}, minima of the reduced
functional (\ref{eq:en}) are also critical points of the full action.
Therefore, our problem reduces to finding solutions of the equation
\begin{eqnarray}
\left(L+\frac{2}{L}\frac{\sin ^2{F(\psi)}}{\sin ^2{\psi
}}\right)\sin ^2{\psi }F''(\psi)
+\left(L+\frac{1}{L}\frac{\sin{2F(\psi)}}{\sin{2\psi}}
F'(\psi)\right)\sin{2\psi
}F'(\psi)\nonumber \\
-\left(L+\frac{1}{L} \frac{\sin ^2{F(\psi)}}{\sin ^2{\psi
}}\right)\sin{2F(\psi)}=0.\label{eqeig:main}\end{eqnarray}
Solutions of this equation were originally analyzed in
\cite{manton86,manton87} (where also a nice geometrical point of
view on the Skyrme model was presented). Full spectrum of these
solutions was studied in \cite{my} and their stability analysis
was carried out in \cite{my2}. A special attention we pay to
energetical stability analysis of the only solution known
analytically, that is, the identity solution $F(\psi)=\psi$ which
we denote by $\ha{1}$. This is the only solution which exist for
$L<\sqrt{2}$ in the topological sector $Q=1$ ($Q$ is an integer,
the topological charge, and is defined here for finite energy
solutions by $F(\pi)-F(0)=Q\pi$). For $L>\sqrt{2}$ and
sufficiently small, another solution exist, whose energy is finite
and tends, as $L\rightarrow\infty$, to the energy of the
flat-space $1$-skyrmion. We denote the solution by $\sk{1}$. This
solution bifurcates from $\ha{1}$ at the critical radius
$L=\sqrt{2}$ and appears due to instability of $\ha{1}$. To
understand this fact, it suffices to analyze the behaviour of the
energy functional $\mathcal{U}[F+\epsilon \xi]$ at $F(\psi)=\psi$,
where $\epsilon$ is a small number and $\xi(\psi)$ are spherically
symmetric perturbations which vanish on boundaries, that is, for
which $\xi(0)=\xi(\pi)=0$. Since $F(\psi)=\psi$ is a solution, the
first variation vanishes $\delta\mathcal{U}[F](\xi)=0$. Therefore,
we examine the second variation which reads
\begin{eqnarray}
\delta^2\mathcal{U}[F](\xi,\xi)=\int\limits_{0}^{\pi}4\pi\sin^2{\psi}
\left\{\left(L+\frac{2}{L}\frac{\sin^2{F}}{\sin^2{\psi}}\right)\xi'^2
+\frac{4}{L}\frac{\sin{2F}}{\sin^2{\psi}}F'\xi
\xi'\right\}\nonumber \\
+\int\limits_{0}^{\pi}4\pi\sin^2{\psi}\left\{\left[\frac{2}{L}\left(1+2\cos{2F}\right)
\frac{\sin^2{F}}{\sin^4{\psi}}
+\frac{2\cos{2F}}{\sin^2{\psi}}\left(L+\frac{F'^2}{L}\right)\right]\xi^2\right\}.
\end{eqnarray}
The general theorem due to Hilbert \cite{hilbert} states that the
consecutive minima of the functional above, at $F(\psi)=\psi$, are
solutions of the equation
$$-(\sin^2{\psi}\xi'(\psi))'+
2\left(1-2\frac{L^2+1}{L^2+2}\sin^2{\psi}\right)\xi(\psi)=
\frac{L}{2+L^2}\sin^2{\psi}\lambda \xi(\psi)
$$
which is the Euler-Lagrange equation
$\delta_{\xi}(\delta^2\mathcal{U}[F](\xi,\xi))=0$. These solutions
span a complete and denumerable linear space of functions $\xi_n$
which are mutually orthogonal with respect to the scalar product
$g(u,v)=4\pi\int_0^{\pi}u(\psi)v(\psi)\sin^2{\psi}\ud{\psi}$. The
$\lambda_n$ are the corresponding eigenvalues given by
$$\lambda_n=\frac{2}{L}(n^2+4n+1)+L(n^2+4n-1), \quad
n=0,1,2,\dots$$ and they are positive for all $L$ with the exception of
$\lambda_0$, which is positive for $L<\sqrt{2}$ and negative for
$L>\sqrt{2}$. The corresponding (unnormalized) eigenfunctions read
$\xi_{2n}(\psi) =\sum_{k=0}^{n} (2k+1)\sin{((2k+1)\psi)}$ and
$\xi_{2n+1}(\psi)=\sum_{k=1}^{n+1} 2k\sin{(2k\psi)}$. Thus, for
$L>\sqrt{2}$, $\ha{1}$ is no longer the absolute minimum of the
energy functional (\ref{eq:en}) in the sector $Q=1$, since then
the perturbation $\xi_0=\sin{\psi}$ decreases energy of $\ha{1}$.
Due to the fact that $\xi_n$ are linear combinations of sine
functions of the argument $k\psi$, the alleged global minimum
$\sk{1}$ can be decomposed into a series of these functions. By
substituting into (\ref{eqeig:main}) one obtains, by formal Taylor
expansion about $L=\sqrt{2}$, that the profile of $\sk{1}$ on the
right of $L=\sqrt{2}$ and $L$ sufficiently small, is given by
$$F(\psi,x)=\psi+x\sin{\psi}+\frac{3}{20}x^2\sin{2\psi}-x^3
\left(\frac{29369}{316800}\sin{\psi}-\frac{11}{480}\sin{3\psi}
\right)+o(x^3), $$ and the corresponding energy reads
$$\frac{\mathcal{U}[F]}{12\pi^2}(x)=\frac{3\sqrt{2}}{4}
\left(1+\frac{11}{180}x^2-\frac{209}{10800}x^4+
\frac{5209}{864000}x^6\right)+O(x^8),\quad x^2=
\sqrt{\frac{60}{11}\left(\frac{L}{\sqrt{2}}-1\right)},$$ which is
less then the energy of $\ha{1}$. The series, if divergent for
large $x$, should have its analytical continuation and reproduce
the energy of the flat space $1$-skyrmion known from the original
Skyrme model. As $x\searrow0$ the profile of $\sk{1}$ smoothly
coalesces with that of $\ha{1}$. Note that
$\lim_{x\searrow0}\partial_xF(\psi,x)=\sin{\psi}$ which is the
mode of instability of $\ha{1}$. It is worth to note also, that
such expansion procedure fails on the left of the critical radius
$L=\sqrt{2}$.

This shows the mechanism how solutions may appear at some critical
parameters of a coupling constant of a theory and not exist if the
coupling is too large. In the case of the $1$-skyrmion this
phenomenon could be analyzed exactly.

\end{document}